\documentclass[floatfix,twocolumn,aps,prl,reprint,superscriptaddress]{revtex4-1}
\usepackage{newfloat}

\usepackage{amsmath}
\usepackage{amssymb}
\usepackage{float}
\usepackage{graphicx}
\usepackage{color,soul}
\usepackage{hyperref}
\usepackage[colorinlistoftodos]{todonotes}
\usepackage{lipsum}
\usepackage{braket}
\usepackage{blindtext}

\usepackage{tablefootnote}

\usepackage[separate-uncertainty=true, multi-part-units=single]{siunitx}

\graphicspath{ {figures/} {figures_supplement/} }

\makeatletter

\@ifundefined{textcolor}{}
{%
 \definecolor{BLACK}{gray}{0}
 \definecolor{WHITE}{gray}{1}
 \definecolor{RED}{rgb}{1,0,0}
 \definecolor{GREEN}{rgb}{0,1,0}
 \definecolor{BLUE}{rgb}{0,0,1}
 \definecolor{CYAN}{cmyk}{1,0,0,0}
 \definecolor{MAGENTA}{cmyk}{0,1,0,0}
 \definecolor{YELLOW}{cmyk}{0,0,1,0}
}
\usepackage{dcolumn}
\usepackage{bm}
\usepackage{float}
\newcolumntype{C}[1]{>{\centering\arraybackslash}p{#1}} 

\usepackage{graphicx}
\makeatother

\begin{document}

\title{Niobium coaxial cavities with internal quality factors exceeding 1.5 billion for circuit quantum electrodynamics}

\author{Andrew E. Oriani}
\thanks{Currently at HRL Laboratories LLC}
 \affiliation{James Franck Institute, University of Chicago, Chicago, Illinois 60637, USA}
\affiliation{Pritzker School of Molecular Engineering, University of Chicago, Chicago, Illinois 60637, USA}

\author{Fang Zhao}
\affiliation{Fermi National Accelerator Laboratory, Batavia, Illinois 60510, USA}

\author{Tanay Roy}
\affiliation{James Franck Institute, University of Chicago, Chicago, Illinois 60637, USA}
\affiliation{Department of Physics, University of Chicago, Chicago, Illinois 60637, USA}

\author{Alexander Anferov}
\affiliation{James Franck Institute, University of Chicago, Chicago, Illinois 60637, USA}
\affiliation{Department of Physics, University of Chicago, Chicago, Illinois 60637, USA}

\author{Kevin He}
\thanks{Currently at HRL Laboratories LLC}
\affiliation{James Franck Institute, University of Chicago, Chicago, Illinois 60637, USA}
\affiliation{Department of Physics, University of Chicago, Chicago, Illinois 60637, USA}

\author{Ankur Agrawal}
\thanks{Currently at AWS Quantum Technologies}
\affiliation{James Franck Institute, University of Chicago, Chicago, Illinois 60637, USA}
\affiliation{Department of Physics, University of Chicago, Chicago, Illinois 60637, USA}

\author{Riju Banerjee}
\affiliation{James Franck Institute, University of Chicago, Chicago, Illinois 60637, USA}
\affiliation{Department of Physics, University of Chicago, Chicago, Illinois 60637, USA}

\author{Srivatsan Chakram}
\affiliation{Department of Physics and Astronomy, Rutgers University, Piscataway, New Jersey 08854, USA}

\author{David I. Schuster}
\affiliation{James Franck Institute, University of Chicago, Chicago, Illinois 60637, USA}
\affiliation{Department of Physics, University of Chicago, Chicago, Illinois 60637, USA}

\date{\today}

\begin{abstract}
 
    Group-V materials such as niobium and tantalum have become popular choices for extending the performance of circuit quantum electrodynamics (cQED) platforms allowing for quantum processors and memories with reduced error rates and more modes. The complex surface chemistry of niobium however makes identifying the main modes of decoherence difficult at millikelvin temperatures and single-photon powers. We use niobium coaxial quarter-wave cavities to study the impact of etch chemistry, prolonged atmospheric exposure, and the significance of cavity conditions prior to and during cooldown, in particular niobium hydride evolution, on single-photon coherence. We demonstrate cavities with quality factors of $Q_{\rm int}\gtrsim 1.4\times10^{9}$ in the single-photon regime, a $15$ fold improvement over aluminum cavities of the same geometry. We rigorously quantify the sensitivity of our fabrication process to various loss mechanisms and demonstrate a $2-4\times$ reduction in the two-level system (TLS) loss tangent and a $3-5\times$ improvement in the residual resistance over traditional BCP etching techniques. Finally, we demonstrate transmon integration and coherent cavity control while maintaining a cavity coherence of \SI{11.3}{ms}. The accessibility of our method, which can easily be replicated in academic-lab settings, and the demonstration of its performance mark an advancement in 3D cQED.

% could use ex: $2$--$4\times$ to get shorter dashes to indicate range, if desired -Kevin
% could put +/- on the 11.4 ms to indicate uncertainty -Kevin

\end{abstract}

\maketitle

\section{\label{sec:level1} Introduction}

Three-dimensional (3D) superconducting cavities, which can preserve long cavity lifetimes while maintaining high-cooperativity when coupled with a transmon qubit, have been used to successfully construct state-of-the-art cQED systems~\cite{reagor2016quantum}. These advances have enabled universal cavity control~\cite{heeres2015cavity, heeres2017implementing} and allowed for demonstrations of quantum error correction~\cite{ofek2016extending, hu2019quantum, campagne2020quantum} and fault tolerance~\cite{reinhold2020error}. These ideas have been extended to multimode systems~\cite{wang2016schrodinger, Chakram2021}, demonstrating two-mode gate operations~\cite{rosenblum2018cnot,gao2019entanglement}, and multimode entanglement~\cite{chakramhe2020}. In addition, 3D cQED systems have also been used for quantum chemistry simulations~\cite{ChrisWang2020}, realizing topological photonic matter \cite{ClaiOwens2021}, and dark matter detection~\cite{dixit2020searching, agrawal2023stimulated},  demonstrating the utility of such 3D systems for applications in quantum information and simulation.

The majority of the cavities used in the aforementioned works have been made of aluminum, thanks to its well-behaved oxide with low dielectric and TLS loss \cite{Reagor2013}, the availability of aluminum in ultra-high purity grades ($>$5N), and its relatively straightforward etching process.  Aluminum can be consistently employed to construct seamless \textrm{$\lambda/4$} co-axial cavities with intrinsic quality factors of \textrm{$>1\times10^{8}$} \cite{kudra2020high}, which can easily be integrated with transmon qubits to realize cQED systems with one millisecond coherence times \cite{reagor2016quantum, kudra2021robust}.

An avenue for increasing the cavity coherence times of 3D cQED systems is to make the cavities out of bulk niobium. Elliptical niobium cavities used for accelerator applications have recently demonstrated single-photon lifetimes of $\tau_{\rm int}>$\SI{2}{s} for a \SI{1.3}{GHz} cavity, with loaded lifetimes of approximately \SI{230}{ms} for a \SI{5.0}{GHz} cavity after reducing TLS loss at the surface \cite{Romenenko2020, Romenenko2017}, showing the potential for such cavities in creating a next-generation of quantum memories. Recent results have also demonstrated qubit integration into elliptical cavities without severe degradation to overall cavity performance \cite{milul2023superconducting}.

In this paper, we study the impact of niobium material properties in the context of operating such cavities at single-photon powers and millikelvin temperatures necessary for cQED. In particular, we delve into the impact of etch chemistry on cavity performance, highlight the significance of air exposure on the TLS loss tangent, and demonstrate only marginal degradation of cavity quality due to hydride evolution when cooling to dilution temperatures. We develop a modified etching process which consistently achieved internal quality factors of $\mathrm{Q_{\rm int}}>1.4\times10^{9}$ at single-photon powers and frequencies of $f_{0}\approx~$\SI{6.5}{GHz}, all while using the aforementioned easy-to-implement 3D coaxial cavity design while keeping the processing steps accessible to small-scale research labs. Finally, we demonstrate transmon integration and coherent control of an 11.3 ms cavity quantum memory.

\begin{figure*}[ht]
   \centering
    \includegraphics[width= 0.8\textwidth]{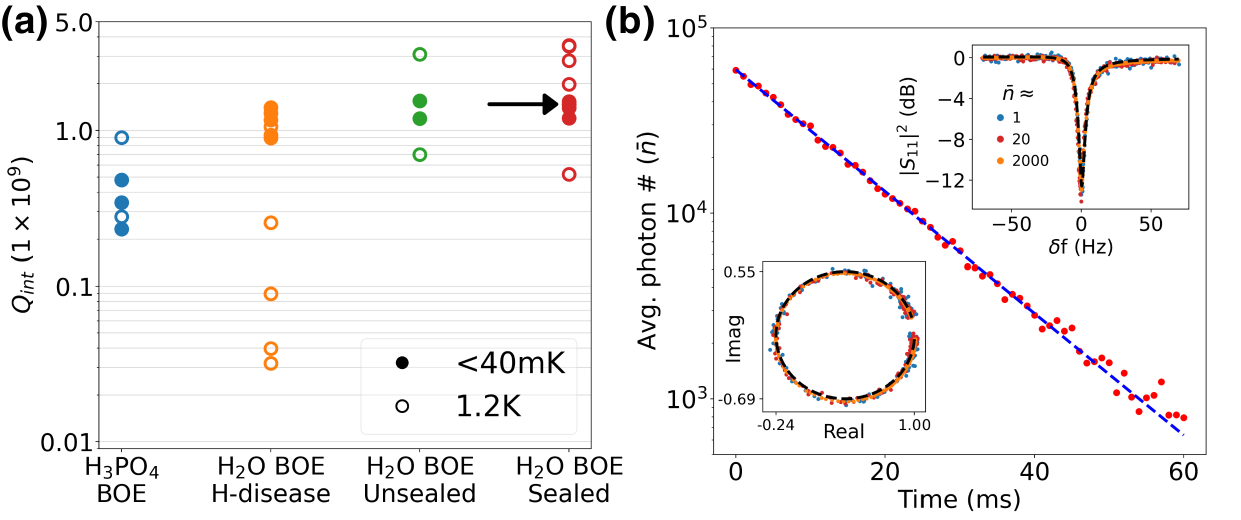}

    \caption{\textbf{Cavity performance survey and cavity spectroscopy} (a) shows a survey of all the cavities measured, organized by etch recipe (H\textsubscript{3}PO\textsubscript{4} and H\textsubscript{2}O), hermetic sealing immediately following etching, and NbH\textsubscript{x} formation (H-disease). Solid circles represent single-photon data at $T<$\SI{40}{mK}, while solid circles represent $\mathrm{Q_{int}}$ data at \SI{1.2}{K}. All cavities etched using the water based buffered oxide etch show 5$\times$ improvement over H\textsubscript{3}PO\textsubscript{4} buffered etch. The mean $\mathrm{Q_{int}}$ for all H\textsubscript{2}O BCP cavities was 1.40$\pm0.128\times10^{9}$. (b) Pulsed-probe spectroscopy of the cavity highlighted in (a) by the arrow showed no power dependence in the cavity linewidth (inset upper right) between $\bar{n}\approx1-2000$ photons. A fit of the data, also presented for the real and imaginary quadratures, give $\mathrm{Q_{\rm int}}=1.47\times10^9$ for all powers. Measurement setup shown in Fig. \ref{SFig 2}.
    }
  	\label{Fig1}
\end{figure*} 

\section{Cavity Design and Preparation}
The coaxial cavity design was chosen due to its intrinsically seamless design and ease of manufacturing. The design is made up of a section of coaxial waveguide that is shorted at one end, leading to a $\lambda/4$ mode. This coaxial section is coupled to a circular waveguide with a cutoff frequency higher than the $\lambda/4$ mode, leading to evanescent loss that is exponentially suppressed along the waveguide’s length. By machining the coaxial section and waveguide out of a single piece of material, we can exclude loss induced by supercurrents crossing a seam, allowing for the more careful study of intrinsic material loss mechanisms \cite{Brecht2015}.

\begin{figure*}[ht]
   \centering
    \includegraphics[width= \textwidth]{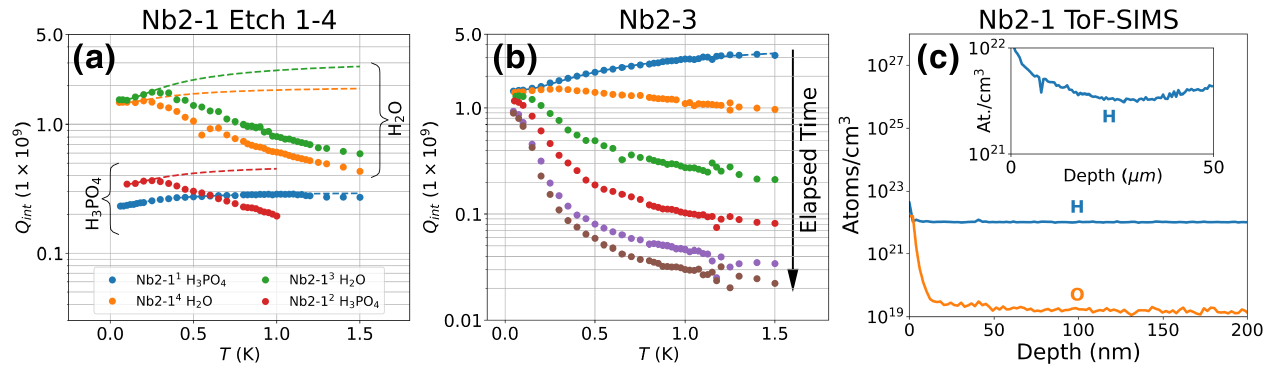}
    \caption{\textbf{Quality factor dependence on etch chemistry and cooldown speed.} (a) Cavity Nb2-1 quality factor comparison between phosphoric acid and water buffered etch. The first two etchings in H\textsubscript{3}PO$_{4}$ showed only mild improvement between successive etchings (\SI{100}{\micro m} each). Successive water buffered etches saw single-photon improvements of 5$\times$. An aborted cooldown between etch 1 and 2 led to a persistent fall-off in $\mathrm{Q_{\rm int}}$ as a function of temperature. This phenomenon is further explored in (b), where a cavity is repeatedly held at \SI{100}{K} for 1 hour before being remeasured. A noticeable decrease in $\mathrm{Q_{\rm int}}$ at high temperature is suggestive of NbH evolution in the bulk, which acts as a magnetic loss mechanism. ToF-SIMS spectroscopy (c) of the Nb2-1$^4$ cavity following the cooldown shown in (a) shows elevated hydrogen concentrations within the London penetration layer extending deep into the bulk, and comparatively low oxygen concentrations after only the first 10~nm.   
    }

  	\label{Fig2}
\end{figure*} 

\par Following etching, rinsing and drying (Appendix \ref{section:cav_etch}), the cavity was clamped in between two copper flanges. The microwave coupler and top flange were hermetically sealed using indium O-rings (see Fig.~\ref{SFig 2}). The cavity was evacuated to a pressure $P<$\SI{1e-5}{mbar} and sealed via a copper pinch-off tube to ensure that atmospheric gases don't condense on the surface and to reduce continued surface oxide formation. The cavity was cooled to $\leq$\SI{45}{mK} in a dilution refrigerator. A detailed discussion of the measurement setup can be found in Appendix \ref{section:meas_setup}.

\section{Temperature Dependence of $\mathrm{Q_{\rm int}}$}

To understand the effect that etch chemistry has on performance, two buffered chemical polish (BCP) recipes were used; one with the traditional 1:1:2 ratio of HF:HNO\textsubscript{3}:H\textsubscript{3}PO\textsubscript{4}, and one with 1:1:2 HF:HNO\textsubscript{3}:H\textsubscript{2}O. The results, shown in Fig.~\ref{Fig1} (a), show cavities etched using H\textsubscript{2}O based BCP had $3-5\times$ improvement in single-photon coherence, and lower two-level system loss and residual resistance when compared with H\textsubscript{3}PO\textsubscript{4}. This was even true for cavities that had previously been etched with H\textsubscript{3}PO\textsubscript{4} BCP before being re-etched using the H\textsubscript{2}O buffered mixture, showing that performance was not intrinsic to a cavity, but a function of etch chemistry. This difference in performance and evidence of the underlying mechanism is discussed in Sec.~\ref{sec:surface chemistry and analysis}.

All cavities etched with H\textsubscript{2}O BCP exhibited performance in excess of $1.3\times10^9$ when cooled directly after etching. Power sweeps, presented in Fig.~\ref{Fig1} (b), showed no noticeable impact on performance from $\bar{n}\approx1-2000$ photons (insets), suggesting weak coupling of defects to the electromagnetic field. The measurements, done near critical coupling ($Q_{\rm int}\approx Q_c$), had loaded coherence times of $\geq$\SI{15}{ms} (Fig.~\ref{Fig1} (b)). 

Achieving such high intrinsic quality factors in Nb cavities requires managing multiple loss channels, notably magnetic loss due to niobium hydride, dielectric loss due to surface oxide, and additional loss from impurities introduced during etching. 

By studying the temperature or power dependence of the internal quality we can extract the two-level system loss tangent ($\delta_{TLS}$), and the residual resistance ($R_s$) simultaneously. The temperature dependence of $Q_{\rm int}$ was used due to the linearity of the cavity at drive strengths of $\geq 2000$ photons. For these measurements the mixing chamber plate was stabilized to a temperature accuracy of $\pm$\SI{5}{\milli K} prior to conducting time-domain spectroscopy. 

One noticeable source of magnetic loss are niobium hydride (NbH\textsubscript{x}) precipitates found near the cavity surface. The species form when interstitial hydrogen form NbH\textsubscript{x} compounds between $50-$\SI{100}{K}. These NbH\textsubscript{x} act like a proximitized superconductor with a lower $T_c$ than the surrounding niobium. Fig.~\ref{Fig2} (a) shows the temperature dependence of cavities that have undergone traditional H\textsubscript{3}PO\textsubscript{4} etching, and that of water-based buffered etch on the same cavity. Although the etching shows improvement in single-photon performance following the H\textsubscript{2}O BCP etch, the temperature-dependent $Q_{\rm int}$ deviates from the first etch, with a faster falloff in $Q_{\rm int}$ at $T\geq$\SI{200}{mK}, suggesting a lowering in $\mathrm{T_c}$.

To systematically study this effect, a new cavity was etched using the H\textsubscript{2}O buffer recipe and cooled. The cavity was measured before and after repeatedly being raised to \SI{100}{K}, and cooled back
down, with 10 hours being spent between $50-$\SI{100}{K} during each cycle.

The results, shown in Fig.~\ref{Fig2} (b), show a monotonic decrease in quality factor after only several temperature cycles. This is congruent with the lowering of $\mathrm{T_c}$ due to NbH\textsubscript{x} formation at the surface due to the presence of hydrogen interstitials introduced during the etching process \cite{Knobloch2003}. Single-photon quality factor decreased by 36$\%$ over the same range. To determine the cause, time-of-flight secondary ion mass spectroscopy (ToF-SIMS) of the Nb2-1 cavity’s center pin, taken post-cooldown and shown in Fig.~\ref{Fig2} (c), shows a large hydrogen background extending deep into the bulk. By comparison, oxygen content becomes vanishingly small beyond $\SI{10}{\nano\meter}$. In SRF cavities this effect has been well characterized, with experiment suggesting cooldown rates in excess of 1K/min needed to avoid NbH\textsubscript{x} formation \cite{Knobloch2003, CZAntoine2003}. By comparison, all cavities in this paper, except for Nb2-1, were initially cooled at rates of $\approx0.3-$\SI{0.5}{K/\minute} between $150-$\SI{75}{K} without variation on performance, even after subsequent cooldowns. While the temperature dependence of $Q_{int}$ is comparable to that of elliptical SRF cavities cooled to dilution temperatures \cite{Romenenko2020}, higher overall residual resistance when compared to cavities that have undergone hydrogen removal still suggests some impact on performance \cite{Ciovati2010}.

\begin{figure*}[ht]
   \centering
    \includegraphics[width= 0.95\textwidth]{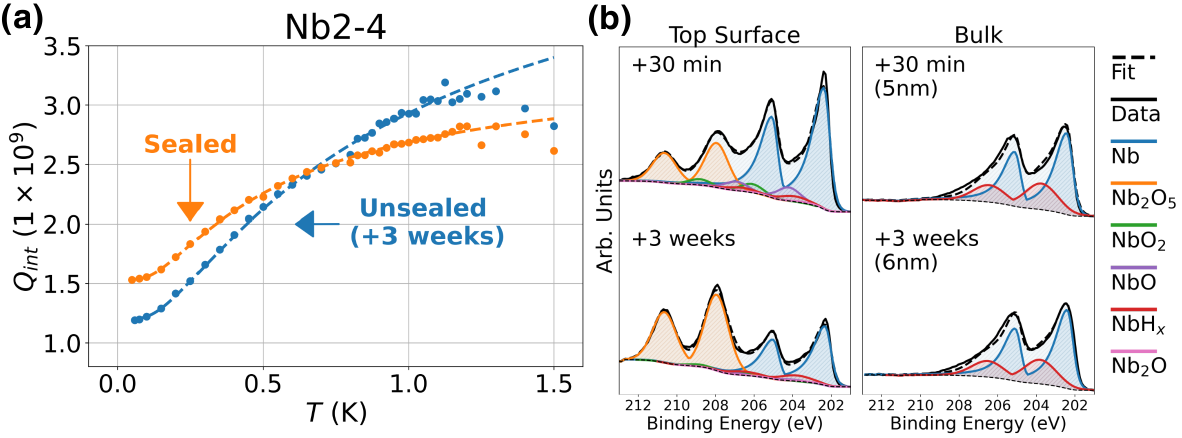 }\caption
    {\textbf{Effect of air exposure time on oxide evolution and cavity performance} (a) temperature dependent cavity performance for cavities hermetically sealed and placed under vacuum following etching when compared to cavities left exposed to air for extended time. The unsealed cavity shows both a 22$\%$ decrease in single photon performance, and 75$\%$ increase in the TLS loss tangent ($\delta_{TLS}$). X-ray photoemission spectroscopy (b) for the cavity directly following a water buffered etch and exposed to air for 22 days, showing the evolution of the oxide. Fitted peaks show an evolution of Nb\textsubscript{2}O\textsubscript{5} in that time, with an extracted increase in oxide thickness of 38$\%$ in that time. Analysis of the sub-surface layer also suggests the continued presence of NbH\textsubscript{x} compounds, corroborating results in Fig.~\ref{Fig2} (b) and (c).
    }
    \label{Fig3}
\end{figure*}

With NbH\textsubscript{x} evolution explored, a third cavity was cooled to better investigate the evolution of the oxide and subsequent two-level system (TLS) loss mechanisms during and after the H\textsubscript{2}O etching process. Niobium is known to host TLS in its oxide layer, which is believed to be caused by the presence of NbO\textsubscript{x} states within a disordered $\approx$\SI{5}{\nano m} thick Nb\textsubscript{2}O\textsubscript{5} surface oxide \cite{Romenenko2017, Premkumar2021, Bafia2024}. To quantify this the extracted temperature-dependent quality factors were fit to the following TLS temperature model \cite{Gao2008,lei2020high, Romenenko2020}:

\begin{equation}
\dfrac{1}{Q_{\rm int}(T)}=\dfrac{1}{Q_{0}}+F_{e}\tan{(\delta_{TLS})}\tanh{\bigg(\alpha\dfrac{\hbar\omega_{0}}{2k_{b}T}\bigg)}
    \label{TLS Eq}
\end{equation} 

Where $F _{e}=t_{\rm ox}S_{e}/\epsilon_{r}$, with $S_{e}$ and $t_{\rm ox}$ being the surface to volume participation ratio of the electric field and surface dielectric thickness respectively. $Q_0=\mu_0\omega_0/R_s S_m$, where $S_m$ is the magnetic surface to volume participation ratio, and $R_s$ is the residual surface resistivity. $S_{e}$ and $S_{m}$ were determined via Ansys HFSS to be \SI{392}{m^{-1}} and \SI{548}{m^{-1}} respectively. The loss tangent product, $F_{e}\tan{(\delta_{TLS})}$ for the H\textsubscript{2}O buffered process ranged from $1.8-5.3\times10^{-10}$ when hermetically sealed within 30 minutes of etch quenching, which is as good or better than state of the art EP etched elliptical cavities prior to additional oxide removal \cite{Romenenko2020}, and $2-4\times$ better than H\textsubscript{3}PO\textsubscript{4} buffered etching. 

To study the long-term oxide evolution and TLS loss, cavity Nb2-4 was etched \SI{100}{\micro \meter} and placed under vacuum within 30 minutes of quenching before being loaded into the refrigerator, versus the same cavity undergoing an identical \SI{150}{\micro \meter} etch and left in a lab dry box, unsealed, for 22 days before being cooled. 

The results, shown in Fig.~\ref{Fig3} (a), show a noticeable effect in both single-photon performance and in the TLS loss tangent. The unsealed cavity exhibits a $\approx75\%$ higher loss tangent product ($F_e\tan{\delta_{TLS}}$) while the single-photon $\mathrm{Q_{\rm int}}$ also decreased by $\approx22\%$. Counterintuitively, the TLS saturated $\mathrm{Q_{\rm int}}$ actually \textit{increased}, suggesting that the additional etching decreasing $\mathrm{R_s}$.

\par To investigate this evolution more closely, the unsealed and exposed Nb2-4 cavity was cut open and its surface analyzed immediately following the cooldown. The oxide thickness and composition was analyzed using X-ray photoelectron spectroscopy (XPS). XPS was done at varying depths using \textit{in-situ} argon ion-sputtering. Etch time versus depth was calibrated against an Nb thin-film deposited on sapphire with an atomic force microscopy (AFM) characterized film thickness (see Appendix ~\ref{section:surface_char}). Following initial characterization, the cavity was re-etched and analyzed within 30 minutes of being quenched.

To understand the relative concentrations of compounds at the surface, the XPS spectra, shown in Fig.~\ref{Fig3} (a), was fit for both freshly etched and unsealed cavities that had been exposed to the atmosphere for prolonged periods of time. The fit spectra reveal four oxidation states and the presence of NbH\textsubscript{x}\cite{Kalboussi2025}, particularly in the underlying Nb-metal surface, which are absent in the Nb thin-film shown in Fig.~\ref{SFig 4} (c), that was etched using fluorine reactive ion etching instead of using an acid based etch. The relative ratios of the Nb metal phase to the Nb\textsubscript{2}O\textsubscript{5}, NbO, NbO\textsubscript{2}, and Nb\textsubscript{2}O suggest a structural change in the oxide. Using techniques outlined in, Appendix \ref{section:oxide_thickness}, these ratios were also used to determine oxide thickness \cite{Premkumar2021, Soda2013}. From the fits, the oxide layer grew by 38$\%$, from \SI{3.6}{\nano \meter} to \SI{4.9}{\nano \meter}. These oxide thicknesses are comparable to previously reported values in unbaked electropolished (EP) cavities \cite{Romenenko2017, Premkumar2021}. Given the linear dependence of the TLS loss-tangent with oxide thickness, this increase does not fully account for the 75$\%$ increase in measured loss tangent product in Fig.~\ref{Fig3} (a). This implies that oxide morphology and composition play a comparable role in the measured loss. 

\begin{figure*}[ht]
   \centering
    \includegraphics[width= 0.95\textwidth]{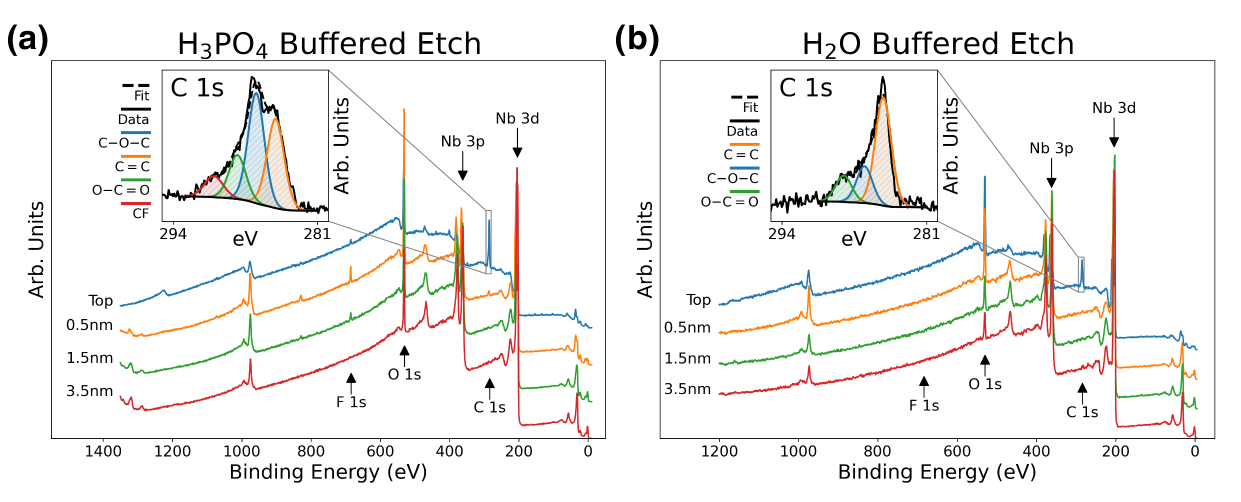}\caption{
   \textbf{Comparison of phosphoric acid and water buffered etch surface chemistry} (a) shows XPS results for the top surface and the immediate underlying oxide layer for a test strip from a phosphoric acid buffer etched cavity (Nb2-1). The inset shows the fitted C1s peak, with deconvolved peaks showing the presence of organic carbon, and fluorine-containing carbon compounds. This coincides with a noticeable F1s peak (684~eV) which persists into the oxide layer. A small C1s peak is also present at \SI{0.5}{\nano m}. A test strip from the water buffered etch of Nb2-1 shows a lack of the F1s and Fkll peaks, and a C1s peak that lacks the asymmetry characteristic of CF\textsubscript{x} compounds.
    }
    \label{Fig4}

\end{figure*}

\section{H\textsubscript{3}PO\textsubscript{4} vs H\textsubscript{2}O buffered etch surface chemistry}
\label{sec:surface chemistry and analysis}

A further question is why the water buffered etch consistently performed better than the more common phosphoric acid buffered etch. During the etch of each cavity, a high-purity ($99.9\%$) Nb test strip was placed in the etch solution, and was measured periodically to determine the etch rate, which varied depending on the temperature and the amount of dissolved NbF products in solution. The test strips from the 2\textsuperscript{nd} H\textsubscript{3}PO\textsubscript{4} buffered etch, and the 3\textsuperscript{rd} H\textsubscript{2}O buffered etch of Nb2-1 (shown in Fig.~\ref{Fig2} (a)) were studied using XPS. 

Survey scans of both test strips, shown in Fig.~\ref{Fig4} (a) and (b), show measurable differences in composition. For the H\textsubscript{3}PO\textsubscript{4} buffered etch, the presence of the flourine 1s peak (\SI{684}{e \volt}) in (a) both at the surface and in the oxide bulk indicates contamination from fluorine-containing products. A careful fitting of the C 1s peak, shown in the Fig.~\ref{Fig4} (a) inset, corroborates this, showing the presence of CF and CF\textsubscript{2} containing compounds. By comparison, the H\textsubscript{2}O buffered etch shows none of these features, with only adventitious organic carbon species at the surface, as indicated in the Fig.~\ref{Fig4} (b) inset.

\section{Discussion}
Surface analysis using XPS and ToF-SIMS shows both the significant introduction of hydrogen to depths of \SI{50}{\micro \meter} and the strong dependence of exposure time to atmosphere, increased oxide thickness, and a subsequent increase in the TLS loss of the resonator. Rapid sealing and evacuation ($<$\SI{90}{min}) seems to keep the TLS loss product at values comparable to electropolished Nb elliptical cavities before any additional mid-T baking to diffuse the oxide \cite{Romenenko2020, Romenenko2017}. This is notable since the geometric participation is $\approx3\times$ greater in the coaxial design than the $\rm TM_{010}$ mode of an elliptical cavity, with a filling fraction $F_e$ that is also $\approx3\times$ greater than that reported for \SI{5}{GHz} TESLA designs \cite{Romenenko2020}. This means that, when scaled, the $R_s$ are within a factor of 3 of the state-of-art Nb elliptical cavities of similar frequencies, and effective TLS loss-tangents are between $2-4\times$ lower than normal EP treated cavities with no additional oxide baking\cite{Romenenko2020}. This, combined with an increase in TLS loss that is greater than what may be expected from increased oxide thickness alone as shown in Fig. ~\ref{Fig3} (a), suggests that H\textsubscript{2}O BCP etched cavities have an oxide with fewer oxygen vacancies than other wet-oxidation techniques \cite{Bafia2024, Grundner1980}. A careful accounting of excess oxygen counts in ToF-SIMs may be one way to rectify this difference between EP and the H\textsubscript{2}O BCP presented here\cite{Bafia2024}. 

% XPS analysis in Fig.~\ref{Fig3} (a) seems to hint at the source of this discrepancy, with the H\textsubscript{2}O BCP etched cavities showing a thin oxide ($<$\SI{4}{nm}) that is largely composed of Nb\textsubscript{2}O\textsubscript{5}. The use of HNO\textsubscript{3}, which is a strong oxidant, may preferentially produce pentoxide over other suboxides, producing a lower-loss dielectric surface.

Another insight was the effect that hydrogen interstitials and NbH\textsubscript{x} formation play on single-photon performance, with NbH\textsubscript{x} formation not leading to a significant decrease in single-photon $\mathrm{Q_{\rm int}}$ at cooldown rates between $0.3-$\SI{0.5}{K/min} over successive cooldowns. Even when hydride formation is induced, $Q_{int}$ decrease by only 36\%. While baking would improve $\mathrm{R_s}$, the above data shows a peak in $\mathrm{Q_{\rm int}}$ at temperatures consistent with cavities that have undergone  $>$\SI{650}{\celsius} baking, suggesting that the cavity $T_c$ is not noticeably affected after a single cooldown cycle \cite{Romenenko2020}.

The final point is the discrepancy between the use of H\textsubscript{3}PO\textsubscript{4} and water as a buffer for the etch and the effect on cavity performance. The consistently higher residual resistance, with no noticeable change in the $T_c$, suggests that the presence of fluorine at the surface, both bonded to carbon but also to Nb, may lead to higher overall loss. Heidler \textit{et. al} \cite{Heidler2021}, who used a nearly identical coaxial design and a similar actively cooled H\textsubscript{3}PO\textsubscript{4} BCP recipe, recorded nearly identical results in TLS and single-photon performance to the highest performing H\textsubscript{3}PO\textsubscript{4} buffer oxide etched (BOE) cavities shown in Fig.~\ref{Fig1} (a).

Previous analysis of electropolished Nb elliptical cavities, using an H\textsubscript{2}SO\textsubscript{4} electrolyte, found the presence of similar CF\textsubscript{x} compounds at the surface, in addition to hydrolyzed niobium flouride species \cite{CHOUHAN2020}.  A correlation was found between the agitation and removal of a viscous product layer at the Nb surface, and poor post-etch rinsing, with the presence of these surface impurities. The closed nature of the coaxial design makes constant agitation and steady-state removal of products from the participating surfaces difficult. By comparison, BCP etching is typically done under constant flow for elliptical cavities \cite{Tereshkin2003, Boffo:2006ap}. During regular pipetting, the viscous layer can be seen as a dark blue (H\textsubscript{2}O BCP) or green (H\textsubscript{3}PO\textsubscript{4} BCP) film that forms at the surface. The difference in color, and the presence of these deleterious species in XPS suggests that the presence of H\textsubscript{3}PO\textsubscript{4} changes the reaction kinetics in a meaningful way, however further investigation into the exact mechanism would need to be undertaken.  

\section{Integration of Transmon Circuit}

While the primary result of this paper was to characterize and reduce surface loss in niobium superconducting cavities through an improved etch chemistry, we were also able to demonstrate the preservation of cavity lifetimes even after integrating a transmon qubit. We demonstrate the integration of a transmon into the coaxial cavity geometry by measuring the single-photon quality factors using a transmon ancilla. We chose to keep the dispersive coupling small (\SI{29}{\kilo Hz}) to limit the decoherence from the inverse Purcell effect via coupling to the lossy transmon. We injected small coherent states into the cavity and observed photon number splitting using the transmon, enabled by sufficiently large transmon coherence ($T_1, T_2 = 60,$\SI{29}{\micro s}), satisfying $\chi > 1/T_2$. By fitting the time dependence of the photon-number-resolved qubit spectrum, we extracted a cavity lifetime of $11.3\pm0.42$ ms at single-photon powers, which is comparable to the loaded ringdown time of the bare cavities ($\approx$12-16 ms). The measured loss is not limited by the participation of the cavity in either the qubit ($\approx$1/300 ms) or the readout resonator ($\approx$1/2 s), while the coupling to the drive pin leads to a limit of  ($\approx$1/25-1/100 ms). Therefore, the measured single-photon loss is a combination of intrinsic cavity loss and additional loss introduced by coupling to the qubit's sapphire substrate. Details of the qubit integration are provided in the Appendix \ref{section:qubit}.

\section{Conclusion}

 In summary, we have introduced a repeatable method to achieve single-photon internal quality factors of $1.5\times10^{9}$ using a compact coaxial cavity design that is both easily manufacturable and readily integrated with superconducting circuits. This achievement represents a $15\times$ improvement over aluminum cavities of the same geometry. We also showed that NbH\textsubscript{x} does not pose a significant threat to single-photon performance during normal cooldown to dilution temperatures. We have also shown that, while surface oxide evolution can degrade cavity performance, TLS loss can be minimized if the time spent between etching and initial cooldown is kept to less than a few hours. Moreover, we've demonstrated that our method surpasses traditional BCP etching techniques in certain metrics, realizing a $2-4\times$ reduction in TLS loss, and a $3-5\times$ improvement in $R_s$ in identical geometries. This implies that the etch chemistry may be more effective in scenarios where removing etching byproducts from the viscous surface layer is challenging. When adjusted for geometry and frequency, our cavities' performance is within a factor of 3 of comparable-frequency Nb elliptical cavities, with limitations set by dielectric loss in the oxide layer. By integrating established oxide removal and N\textsubscript{2} doping methods, we project at minimum a tripling in single-photon performance. Finally, we achieve transmon integration while maintaining high cavity and transmon lifetime and low thermal populations, and a cavity lifetime of 11.3 ms cavity quantum memory. 
 
 Using the techniques outlined here, we hope to enable higher performance quantum memories, and make niobium cavity manufacturing accessible to a wider scientific audience, while also illuminating the effects of niobium surface chemistry on superconducting RF performance at millikelvin temperature and single-photon powers.

\section{Acknowledgments}
\begin{acknowledgments}

The authors would like to acknowledge Alexander Filatov for his assistance in XPS analysis and Aaron Chou for his programmatic support of this project. ToF-SIMS measurements were done with assistance from Eurofins EAG Laboratories. We would also like to acknowledge Justin Jureller and Luigi Mazzenga for technical assistance throughout. This work was supported by the Samsung Advanced Institute of Technology Global Research Partnership. This work is also supported by the U.S. Department of Energy Office of Science National Quantum Information Science Research Centers as part of the Q-NEXT center. Material fabrication and characterization was partially supported by the University of Chicago Materials Research Science and Engineering Center, which is funded by the National Science Foundation under award number DMR-2011854. Transmon qubit devices were fabricated in the Pritzker Nanofabrication Facility at the University of Chicago, which receives support from Soft and Hybrid Nanotechnology Experimental (SHyNE) Resource (NSF ECCS-2025633), a node of the National Science Foundation’s National Nanotechnology Coordinated Infrastructure. This document was prepared using the resources of the Fermi National Accelerator Laboratory (Fermilab), a U.S. Department of Energy, Office of Science, Office of High Energy Physics HEP User Facility. Fermilab is managed by Fermi Research Alliance, LLC (FRA), acting under Contract No. DE-AC02-07CH11359.

\end{acknowledgments}

\appendix
\setcounter{figure}{0}
\renewcommand{\thefigure}{S\arabic{figure}}

\begin{figure*}[!ht]
   \centering
    \includegraphics[width= 1.0\textwidth]{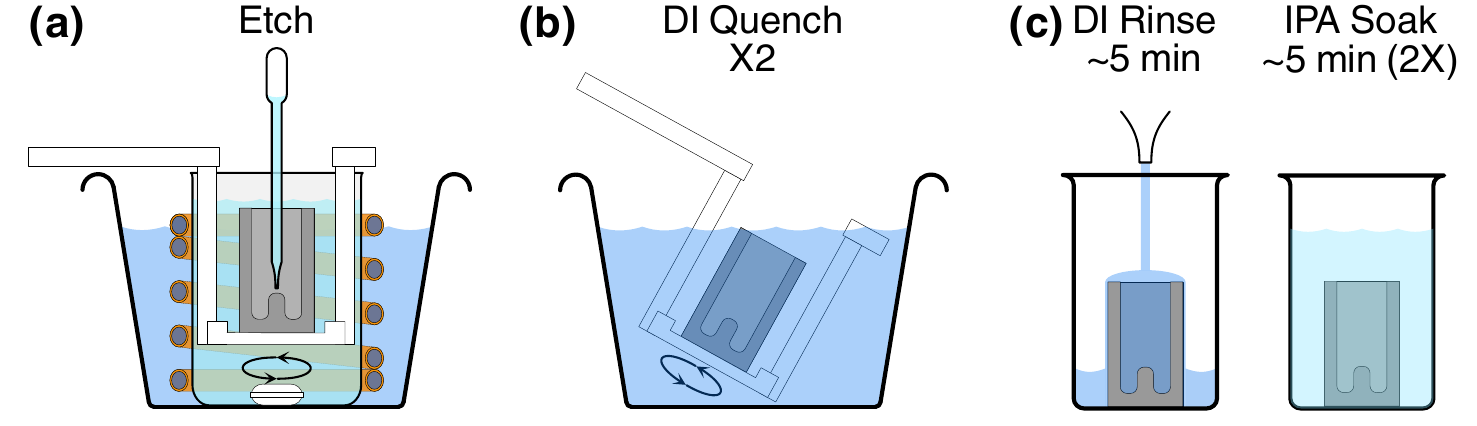}

    \caption{
    \textbf{Etching process steps}: A depiction of the primary etching steps and layout of components. (a) shows a cartoon of the etch setup in cutaway, depicting the copper cooling coil surrounding the primary teflon etch containment. A table of pertinent etching and quenching parameters used in this paper are outlined in Table \ref{Table 1}. (b) depicts the quenching process and information about the quench step in the table below. (c) depicts final DI rinsing and the solvent exchange process used for drying. 
    }
  	\label{SFig 1}
\end{figure*}

\section{Cavity Manufacturing and Etching}

\label{section:cav_etch}
\par The cavity was machined out of a piece of pure niobium metal with a residual resistance ratio (RRR) of $\ge300$ which conformed to ASTM B393 R04220 standards and was purchased from Ningxia Nonferrous Metals. For more even material removal during etching, sharp edges and corners of the co-axial design were radiused for smooth transitions. The cavity was initially degreased using an Alconox$^{\copyright}$ and DI-water mixture at temperatures $>$\SI{50}{\celsius} under sonication. Following this, the cavity was thoroughly rinsed in DI-water and dried using an isopropyl alcohol (IPA) solvent exchange and dry nitrogen. A toluene-acetone-methanol-IPA cleaning was done using semiconductor (semi) grade solvents under sonication for $\geq$\SI{5}{\min} at each step, before being dried under high-purity nitrogen that had been passed through a \SI{0.2}{\micro \meter} filter before being bagged in a ISO100 clean-bag. All cavities were stored in a desiccated dry box after cleaning and prior to etching.

\begin{table}[!ht]
\begin{tabular}{|c|c|}

      \multicolumn{2}{c}{Etching Parameters}\\
      \hline\hline
        % \rowcolor{Rectcolor}
        Etch & 1:1:2 (HF:HNO\textsubscript{3}:H\textsubscript{2}O)$^{\dagger}$  \\ % <--
         \hline
         Etch depth ($\mu$m) & $100-150$\\
         \hline
         Etched area ($cm^{2}$) & 70.2\\
         \hline
         Time (min) & $60-120^{\ddagger}$\\
         \hline
         Temp ($^{\circ}C$) & $<10$\\
         \hline
         Etch vol. (mL) & $\sim500$\\
         \hline
         Stir-rate (min$^{-1}$) & $1250$\\
         \hline
         Pipette Interval (min) & $5-15$\\
         \hline\hline
         \multicolumn{2}{c}{Quenching and Rinse Parameters}\\
         \hline\hline
         Water resistivity ($\Omega~ cm$) & $\geq18.6\times 10^{6}$ \\ % <--
         \hline
         Water TOC (PPB) & $<$10 \\
         \hline
         Quench water vol ($L$) & $\sim2$\\
         \hline
         Quench time, 1\textsuperscript{st} step (min) & $\sim1$\\
         \hline
         Quench time, 2\textsuperscript{nd} step (min) & $\sim5$\\
         \hline
         Temp ($^{\circ}C$) & $22$\\
         \hline
         DI rinse time (min) & $\sim5$\\
         \hline
         Drying solvent & IPA (semi-grade)\\
         \hline
         Solvent exchange time (min) & $\sim 5$ \\
         \hline
         Drying gas & N\textsubscript{2}$^{\S}$\\
         \hline
        % \multicolumn{2}{l}{\footnotesize $^{\dagger}$ Sold as RSE1:1:0:2H\textsubscript{2}O from Transene Co. Inc. ,$^{\ddagger}$ calibrated \textit{in-situ} for specific etch-rate using Nb test-strip, $^{*}$  determined based on etch depth and  surface-area\par}

     \end{tabular}
     \caption{\textbf{Table of etching and quencing parameters:} $\dagger$ Sold as RSE1:1:0:2H\textsubscript{2}O from Transene Co. Inc. $\ddagger$ calibrated \textit{in-situ} for specific etch-rate using Nb  test-strip. $\S$ Oil-free, two-stage filter/drying, $0.2\mu$m filter
     }
      \label{Table 1}
    \end{table}

The top $100-$\SI{200}{\micro \meter} of the surface is etched away to remove the damaged layer that is created from machining processes, and to remove any contaminants that may be at the surface. Buffered chemical polish (BCP) etching, works by first oxidizing the surface before the oxide is etched by hydrofluoric acid (HF). Here the reaction kinetics are primarily dependent on the relative ratios of oxidizer (HNO\textsubscript{3}) to HF, with a buffer added to slow the reaction rate. In addition to a buffer, kinetics are also dependent on the etch temperature and the relative concentration of products in the solution. 

The addition of an H\textsubscript{3}PO\textsubscript{4} or H\textsubscript{2}SO\textsubscript{3} buffer has shown reduced etch rate and improved polishing action \cite{Kinter1970, Uzel1983, antoine1999morphological}. The limiting action of the buffer is believed to be due both to kinetics, but also the presence of a viscous product rich layer at the Nb surface \cite{ASPART200417}. For lower relative concentrations of buffer, the polishing action is reduced, with selective etching occurring at the grain boundaries \cite{Uzel1983}. For the commonly used 1:1:2 mixture of HF:HNO\textsubscript{3}:H\textsubscript{3}PO\textsubscript{4} buffered chemical polished solution at an etching temperature of $\gtrsim10^{\circ}$C, $\approx480-900$W of power is dissipated per \SI{1}{\meter^2} of Nb etched at a rate of $\approx$1\textrm{$\mu$m/min} \cite{Tereshkin2003, kelly2017surface}. Another important factor in final etch quality is the amount of dissolved NbF\textsubscript{5} precipitate in the etch solution, with conventional wisdom being $<20$g/l of dissolved Nb\textsuperscript{5+} ions in solution, with our recipe designed to have $<10$g/l for a $\approx100~\mu$m etch \cite{Tereshkin2003, kelly2017surface}. 

The etch recipe used in this paper diverges from the traditional BCP procedure, using a 1:1:2 HF:HNO\textsubscript{3}:H\textsubscript{2}O etch chemistry. The smaller surface area of our cavity, as listed in Table \ref{Table 1}, means that our dissipated power is \SI{10}{\watt} for even our highest etching rates. This, combined with the smaller volume of etchant required, means that an unbuffered etch can be used with minimal degradation in etch quality while keeping the reaction kinetics in a safe and manageable regime. In addition, the closed nature of the cavity geometry, with poor acid circulation, means that higher viscosity BCP mixtures can greatly affect the etch performance at the surface.

 The main etching steps of the cavity are depicted in Fig.~\ref{SFig 1} Cav etch. A custom Teflon$^{\copyright}$ PTFE holder was built to facilitate rapid and safe transfer of the cavity between etching step (depicted in Fig.~\ref{SFig 1}). PCTFE/PVDF screws were used to attach the cavity to the holder. The main etching was done in a \SI{600}{\milli \liter} PTFE beaker. The beaker was placed inside of a water bath, and surrounded by a copper coil attached to a \SI{400}{\watt} thermoelectric cooler which circulates a water/glycol mixture through the coil. The whole water bath was set on a magnetic stirring plate to circulate the primary acid volume. A pure niobium test strip is placed into the etch container alongside the cavity. The strip is periodically removed, rinsed, and thickness measured using a micrometer to check etch-rate. Fig.~\ref{SFig 1} (a) provides a diagram of the aforementioned etch setup. To circulate the acid inside the cavity volume, the etchant is manually exchanged using a clean polypropylene pipette at $5-15$min intervals. Other pertinent etching parameters are listed in the tables of Fig.~\ref{SFig 1} (a$-$c).  
 
 To prevent unwanted reaction products from adhering to the surface and leaving residue, the etchant is manually exchanged via a pipette for the final minute of etching, before evacuating the BCP solution and dunking it into a (UHP/Type I) DI water bath while vigorously agitated for $\approx1$min, quenching the etching reaction and reducing the chance of dissolved Nb salts and fluorinated compounds from depositing onto the surface. This step is done twice to further dilute any residual etching solution. Pertinent parameters of this step are presented diagrammatically and in a table in Fig.~\ref{SFig 1} (b), with water purity tightly controlled to reduce the introduction of dissolved ions onto the surface, with measured parameters listed in detail in both Fig.~\ref{SFig 1} (a) and (b).

 \begin{figure*}[ht]
   \centering
    \includegraphics[width= 0.9\textwidth]{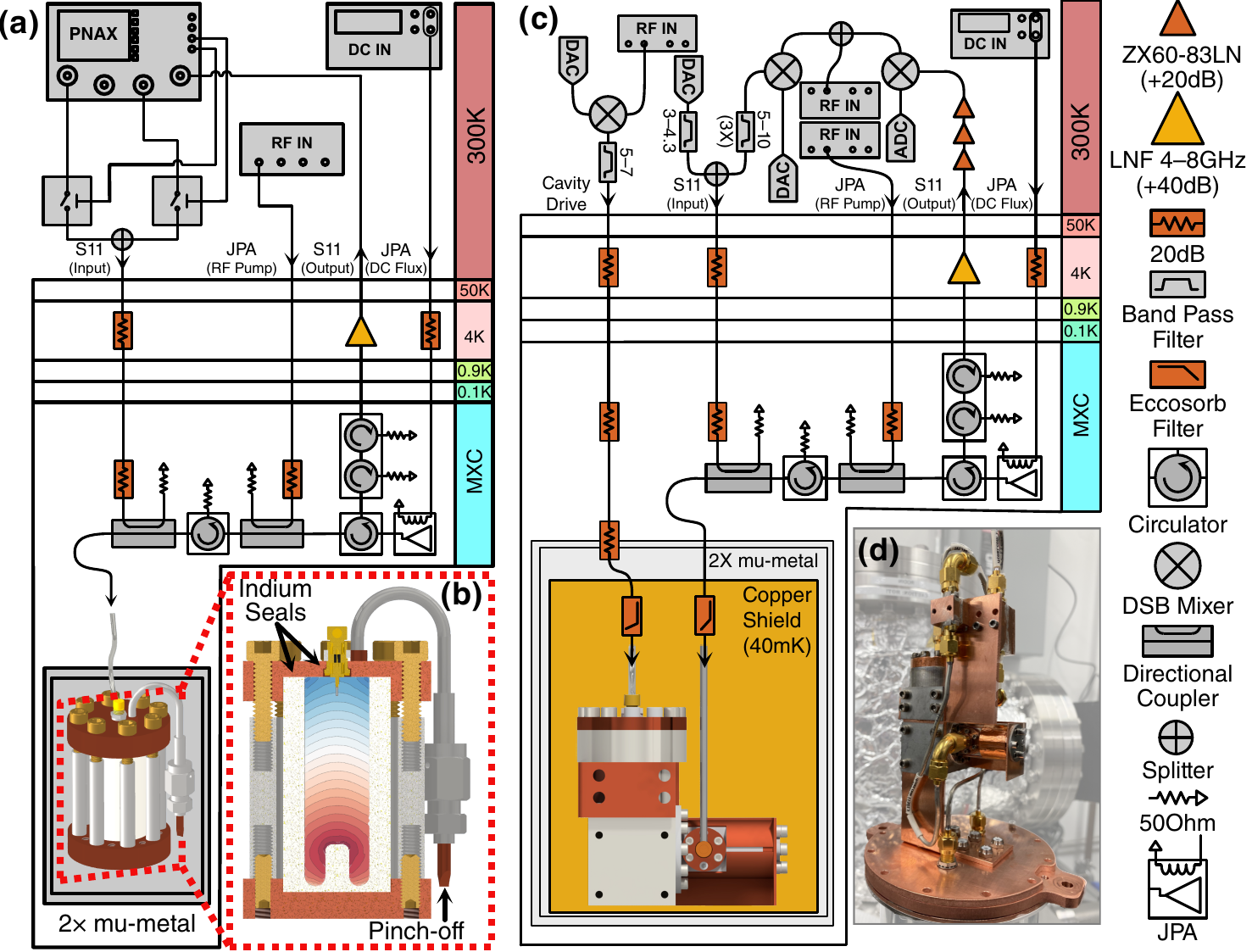}

    \caption{\textbf{Measurement Setup}: (a) Room temperature and cryogenic measurement setup used to do spectroscopy and time-domain measurements of the bare storage cavity. A Keysight PNAX was used to do both CW spectroscopy and time domain measurements as shown in Fig.~\ref{Fig1} (b) Shows a cross-section of the hermetically sealed cavity, highlighting the location of indium seals at the top of the cavity and coupler, along with the integrated pinch-off tube for hermetically sealing the cavity following vacuum pumping of the cavity. (c) shows the modified room temperature and cryogenic measurement setup used to take data on the integrated cavity-qubit system. This includes the addition of a separate coherent drive to the storage cavity and the use of an Xilinx ZCU216 RFSoC in place of the PNAX. (d) is a picture showing the complete storage cavity, Eccosorb filtering, and integrated readout cavity after hermetic sealing. A detailed cutaway of the hermetic readout cavity is shown in Fig.~\ref{Fig4} (a).
    }
  \label{SFig 2}
\end{figure*}

 Following this initial quenching step, the cavity is rinsed under a gravity fed stream of UHP Type-I water for $\approx$\SI{5}{\minute} to remove any stubborn contaminants. Following this the cavity undergoes a two-step solvent exchange process. Here we use semi-grade isopropol alcohol (IPA) to displace water, with each soak being $\approx$\SI{5}{\minute}. Following this, the cavities are dried under high pressure dry nitrogen which has undergone two-stage filtration, with the final stage being a  \SI{0.2}{\micro \meter} filter. 
 
 Following drying the cavity is sealed. An indium ring is sandwiched between the top of the cavity and the cap. A 3mm stainless steel tube is brazed into the copper cap. A Swagelok VCR$^{\copyright}$ fitting is brazed at the other end. A second VCR fitting is attached to a copper pinch-off tube, allowing for easy replacement and reuse of the cap. A Corning-Gilbert GPO (SMP) hermetic connector (Part$\#$ 0119-783-1) is sealed into the top of the cap via a second indium o-ring, and allows for microwave feedthrough. An antenna is made on the vacuum-side of the feedthrough to couple RF power into the cavity. Because of the evanescent decay of the field energy, the coupling quality factor $Q_{ext}$ is exponentially sensitive to this antenna distance from the cavity center pin. The cavity is left under vacuum for $\approx$\SI{24}{\hour} before being sealed via a hydraulic pinch-off tool and placed into the fridge. A diagram of the sealed cavity with pinch-off tube is shown in Fig.~\ref{SFig 2} (b).

\begin{figure*}[ht]
   \centering
    \includegraphics[width= 1.0\textwidth]{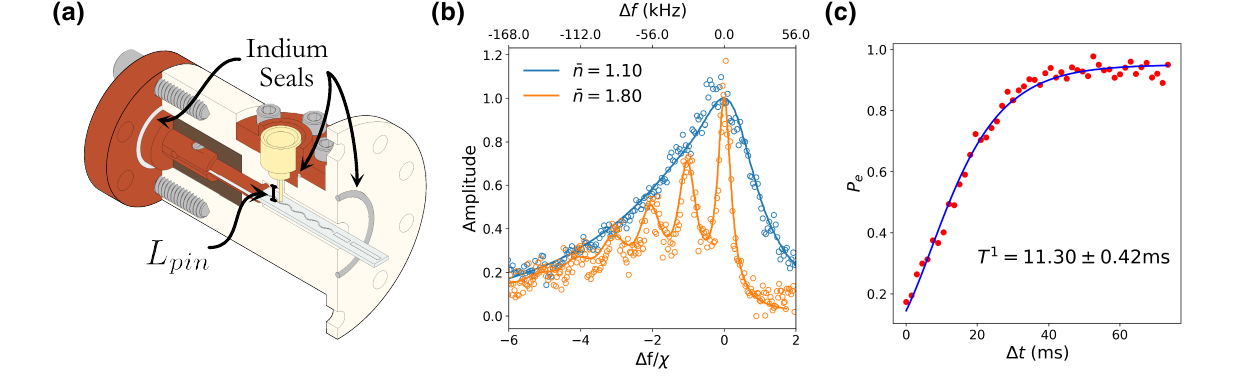}
    \caption{\textbf{Qubit-Cavity integration and characterization} (a) a cutaway of the device showing the qubit mount, chip, and readout couplers, showing the monolithic clamp used to affix the chip and provide thermalization. An indium washer was used at the front and rear of the mount to hermetically seal the cavity. A hermetic GPO connector with a pin extending over the stripline readout resonator was used for qubit readout and manipulation. (b) A plot showing qubit spectroscopy in the presence of photons in the storage cavity, comparing two cases: (1) where the weak dispersive interaction is resolved (orange), and (2) where the interaction is not resolved ($\chi < \kappa$). The only change in the experimental conditions arose from modifying the coupling Q to the readout resonator, and the resulting change in $T_2$ from a reduction in photon shot noise dephasing ($Q_c = 2400 \rightarrow 20000)$. Both data are fit to the functional form obtained for a coherent state, accounting for photon number resolved qubit peaks and including qubit and photon-number-dependent cavity decay~\cite{gambetta2006qubit, schuster2007resolving}. Both fits result in a $\chi\approx$~\SI{28}{kHz}.(c) Qubit spectroscopy was fit to extract the population vacuum state, which was performed as a function of the delay between the cavity drive and qubit spectroscopy pulse to extract the T1 of the cavity. }

    \label{SFig 3}

\end{figure*}

\section{Measurement Setup}

\label{section:meas_setup}

The cavity was measured in reflection ($S_{11}$). The input line uses discrete attenuation at 4K and the mixing chamber plate (MXC) of the dilution refrigerator, with an additional \SI{20}{\decibel} of attenuation being added at the insertion port of a Quantum Microwave cryogenic directional coupler, giving at total of $\approx$\SI{60}{\decibel} of cryogenic attenuation. The reflected signal is sent to a Josephson-parametric amplifier (JPA) through an additional circulator and second directional coupler, which is used to couple the RF pump power into the JPA. Niobium titanium (NbTi) lines then pass the signal to 4K where they are amplified via a $4-$\SI{8}{\giga \hertz} HEMT amplifier~(Low Noise Factory model $\#$ LNC4-8C).

The cavity is measured via a Keysight PNA-X N5242A network analyzer. The time-domain setup is shown in Fig.~\ref{SFig 2} (a). A sequence of pulses generated internally by the PNA-X is used to trigger high-speed (\SI{20}{ns} switching time) reflective SPDT RF switches (Mini-circuits model $\#$ ZASW-2-50DRA+) which pulse the PNAX outputs with the ADC triggered internally. Ringdown measurements are performed by pulsing on the cavity drive and sweeping a delay before turning on the ADC. The JPA RF pump power is provided via a Signalcore SC5510A RF synthesizer, while the DC flux was provided by an external NbTi coil, with the bias current provided by a Yokogawa GS200 precision current source. A picture and cutaway of the hermetically sealed coaxial cavity is shown in Fig.~\ref{SFig 2} (d), highlighting the location of indium seals, the copper pump-out and pinch-off tube, and the coupler location. 

For integrated qubit measurements a different time-domain setup was used, and is shown in Fig.~\ref{SFig 2} (c). An additional cavity drive line was added, while the cryogenic RF drive and readout hardware shown in Fig.~\ref{SFig 2} (a) was moved to the stripline readout cavity, shown in Fig.~\ref{SFig 3} (a). Coherent cavity, qubit, and readout pulses were generated using a Xilinx ZCU216 RFSoC with the QICK firmware \cite{QICK2022}. The signals were upconverted using double-sideband mixing (Mini-circuits model $\#$ ZX05-83-S+), following the readout topology used in Ref. \cite{QICK2022}. Additional Eccosorb filtering was added to the cavity and readout drive lines to reduce readout cavity and qubit excited state populations and resulting shot-noise dephasing of the qubit and the storage cavity, respectively. Fig.~\ref{SFig 2} (d) shows a picture of the mounted cavity-qubit system attached to the thermalization brackets. The entire device, along with the eccosorb filters, were placed inside of a copper-lined double-layer Mu-metal shield to precent errant IR impingement.

\section{Transmon Integration and Measurements}
\label{section:qubit}

To keep cavity performance degradation to a minimum, the qubit chip---with integrated stripline readout---was hermetically sealed inside of an evanescent waveguide that was inserted into the side of the cavity, as depicted in Fig.~\ref{SFig 3} (a). Thermalization of the qubit was done via a novel one-piece mount that held the qubit between two prongs made by cutting a $500~\mu$m slit into a copper post. A bronze-filled PTFE collet was slid over the prongs of the clamp to provide increased clamping force while cold thanks to its high coefficient of thermal expansion. This simplified design made sealing the coupler, qubit mount, and waveguide section to the cavity far easier, with indium o-rings used to create a reliable hermetic connection.

The readout was probed in reflection by coupling a GPO connector to a stripline resonator at $\omega_r=2\pi\times$\SI{7.4}{GHz}. The qubit frequency was set at $\omega_q=2\pi\times$\SI{3.75}{GHz} with $\chi_{rq}\approx$\SI{1}{MHz}. To maintain high cavity coherence we drastically reduce its coupling to the transmon in order to mitigate errors inherited from the lossy qubit. These errors include the inverse-Purcell effect arising from the participation of the transmon in the cavity mode ($\Gamma_{c} = \frac{g^2}{\Delta^2} \Gamma_q$) and cavity-dephasing from excitations of the transmon. We set the coupling to achieve a dispersive shift of $\chi_{sq}=$\SI{28}{kHz}, resulting in a Purcell limited lifetime of $T_{p}\geq500$ ms assuming a qubit $T_1\approx~100~\mathrm{\mu}$s. 

With such a small dispersive shift, resolving the number-splitting of the storage resonator required a narrow enough qubit linewidth ($1/T_2$). We also used long spectroscopy pulses ($T_{q}\geq T_2$). The $T_2$ of the transmon was limited by photon-shot-noise dephasing, arising from thermal photons in the readout cavity. We reduced the dephasing from photon-shot-noise by increasing the coupling Q of the readout cavity. Fig.~\ref{SFig 3} (b) shows qubit spectroscopy in the presence of a cavity drive for two different qubit linewidths, achieved by changing the readout coupling Q. Both spectra were taken with cavity displacement drives corresponding to $\bar{n}\approx1$ photon, and fit using the same dispersive shift but different qubit $T_2$s. The qubit linewidths were obtained by first fitting the $\ket{0}$-photon qubit peak, with peaks corresponding to higher photon numbers having an additional photon-number-dependent broadening \cite{gambetta2006qubit}. The photon-number states were resolved despite the weak dispersive coupling with qubit coherence times of $T_1, T_2 = 60, 29 \mu s$.

Once the cavity displacement was calibrated, the storage cavity $T_1$ in the single-photon regime was obtained by displacing the cavity and measuring the amplitude of the $\ket{0}$ qubit peak following different delay times, as shown in Fig.~\ref{Fig1} (c), resulting in a cavity $T_1=$\SI{11.3}{ms}.  

\section{Surface Characterization}
\label{section:surface_char}

\begin{figure*}[ht]
   \centering
    \includegraphics[width= 1\textwidth]{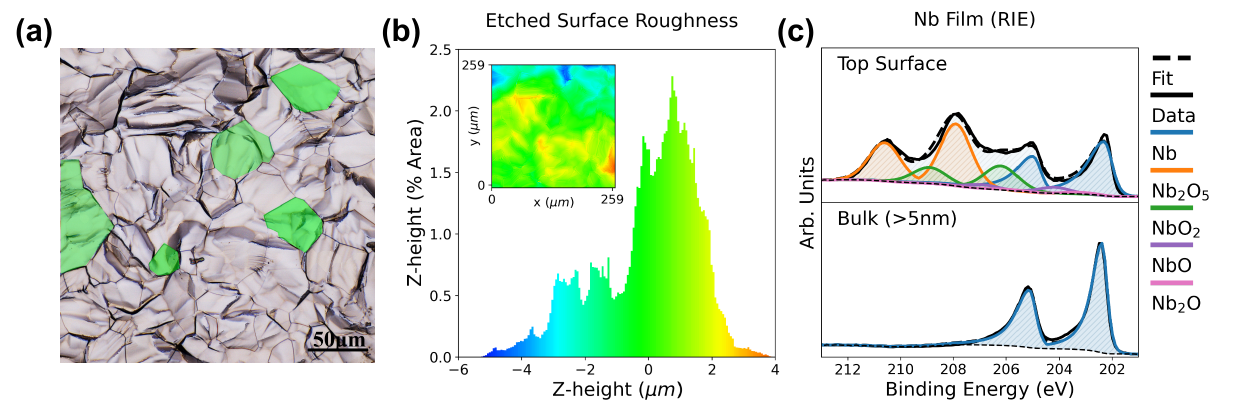}

    \caption{
    \textbf{Surface roughness and Nb film characterization}: (b) A high contrast image using a white-light confocal microscope of a post-etched surface, showing clearly delineated crystal domain edges, with some domains highlighted in green. (b) The distribution of the surface roughness of the patch shown in (a) is plotted as a percentage of the total area, with the surface roughness map shown as an inset. The data gives a surface roughness RMS value of $1-2\mu${m} across the top of the center pin. (c) The XPS spectra of a thin-film Nb sample that was patterned and etched using reactive-ion etching (RIE). This sample was used to calibrate the argon ion sputtering rate and to act as a control for the Nb spectra shown in Fig.~\ref{Fig3} (b). The sample had been exposed to air for an extended period of time ($>$1 month) to allow for the oxide to fully stabilize, with the surface spectra of the Nb film showing similarities to the spectra of the cavity, following prolonged exposure to air. By using reactive ion etching, we removed any possibility of hydrogen diffusion into the bulk, allowing for the collection of a clean Nb 3d spectra. By comparison, the Nb 3d bulk data shown in Fig.~\ref{Fig3} (b) shows a distinct lack of definition due to the formation of NbH\textsubscript{x} compounds. 
    }
  	\label{SFig 4}
\end{figure*}

We performed measurements of surface roughness, as well as the composition of lossy dielectric and conductive compounds within the London penetration depth of the niobium. The surface roughness measurements were performed using an Olympus OLS5000 laser confocal microscope with a 50x objective and a $405$nm scanning light source capable of \SI{1}{nm} $z$-resolution. Fig.~\ref{SFig 4} (a) shows a white-light image of the center pin surface following etching, with green used to highlight several crystal domains. The material used showed a fairly fine-grain crystal structure, with delineated domains ranging in size from $10-50~\mu$m. A 3D scan of this region gave an arithmetic surface roughness of $R_a\approx2~\mu$m. Several other regions studied showed surface roughness ranging from $1-2~\mu$m, which is comparable to values seen in the literature for BCP-treated polycrystalline Nb cavities, and at least an order of magnitude higher than EP etched cavities~\cite{antoine1999morphological} over the same surface area. Fig.~\ref{SFig 4} (b) provides a color $z$-height map of (a) (inset), with a distribution of these deviations as a percentage of area, showing a peak centered at $\approx1.6~\mu$m of surface roughness.

The compositions of both surface oxides and the underlying niobium metal were determined using X-ray photoemission spectroscopy (XPS). Analysis was performed on a Kratos Axis Nova spectrometer using a monochromatic Al K$\alpha$ source (1486.6~eV). Nb 3d, C 1s, and O 1s high-resolution spectra were collected using an analysis area of $0.3\times0.7$~mm and 20~eV pass energy with the step size of 100~meV. The survey spectra were collected using 100~eV pass energy. Charge neutralization was performed using a co-axial, low energy ($\approx$ 0.1~eV) electron flood source to avoid shifts in the recovered binding energy. The C 1s peak of adventitious carbon was set at 284.8~eV to compensate for any remaining charge-induced shifts. Deconvolution of the high-resolution XPS spectra was performed in CasaXPS software using Lorentzian asymmetric curves for the Nb metal phase, symmetric Gaussian-Lorentzian curves for all other elements’ fitting, and a Shirley fitting for the background. The Nb 3d region consists of the two 3d\textsubscript{5/2} and 3d\textsubscript{3/2} spin-orbit split components. The peak area ratio of 3d\textsubscript{5/2} to 3d\textsubscript{3/2} was fixed to 3:2. The Nb 3d region was fit using six doublet components (Nb metal, Nb\textsubscript{2}O\textsubscript{5}, NbO\textsubscript{2}, NbO, Nb\textsubscript{2}O, NbH\textsubscript{x}) of 3d\textsubscript{5/2} and 3d\textsubscript{3/2} components for each sample. The energy splitting of each doublet component is 2.7~eV. 

\onecolumngrid

\begin{table}[ht]
    \centering
    \begin{tabular}{ |C{3cm}|C{3cm}|C{3.5cm}|C{1cm}|C{1cm}|C{1cm}|C{1cm}|C{1cm}|C{1cm}|  }
    \hline
     \setlength{\tabcolsep}{2.5pt}
     \renewcommand*{\arraystretch}{1.25}
     \textbf{Depth} & \textbf{Sample} & \textbf{Parameters} & \textbf{Nb} & \textbf{Nb$_2$O$_5$ 3d$_{5/2}$} & \textbf{NbO$_2$ 3d$_{5/2}$} & \textbf{NbO 3d$_{5/2}$} & \textbf{Nb$_2$O 3d$_{5/2}$} & \textbf{NbH$_x$ 3d$_{5/2}$} \\
     \hline
     \hline
                 & Nb2-4          & Binding Energy (eV)& 202.3 & 207.95 & 206.1 & 204.1 & 203.2  & 203.89 \\ \cline{3-3}
                 & +22 days       & FWHM (eV)          & 0.75  & 1.45   & 1.5   & 1.5   & 1.3    & 1.8    \\ \cline{3-3}
                 &                & Atomic Comp. ($\%$)& 38.1  & 52.59  & 2.2   & 0.98  & 0.54   & 5.47   \\ \cline{2-9}
                 & Nb2-4          & Binding Energy (eV)& 202.35& 207.95 & 206.1 & 204.2 & 203.2  & 203.89 \\ \cline{3-3}
         Surface & +30 min        & FWHM (eV)          & 0.75  & 1.45   & 1.50  & 1.5   & 1.3    & 1.8   \\ \cline{3-3}
                 &                & Atomic Comp. ($\%$)& 60.55 & 23.78  & 4.37  & 7.05  & 0.62   & 2.58  \\ \cline{2-9}
                 & Nb             & Binding Energy (eV)& 202.27& 207.9  & 206.2 & 204.2 & 203.2  & ---    \\ \cline{3-3}
                 & Film           & FWHM (eV)          & 0.75  &  1.4  &  1.50  &  1.40 &  1.40  & ---    \\ \cline{3-3}
                 &                & Atomic Comp. ($\%$)& 35.87 &  41.16 & 17.65 &  4.13 &  1.15  & ---    \\ \hline\hline
                 & Nb2-4          & Binding Energy (eV)& 202.4 &  ---   &  ---  &  ---  &  ---   & 203.90 \\ \cline{3-3}
                 & +22 days       & FWHM (eV)          & 0.75  &  ---   &  ---  &  ---  &  ---   & 2.0    \\ \cline{3-3}
                 &                & Atomic Comp ($\%$) & 68.8  &  ---   &  ---  &  ---  &  ---   & 31.2  \\ \cline{2-9}
                 & Nb2-4          & Binding Energy (eV)& 202.4&  ---   &  ---  &  ---  &  ---    & 203.9 \\ \cline{3-3}
     +35nm depth & +30 min        & FWHM (eV)          & 0.75  &  ---   &  ---  &  ---  &  ---   & 2.0    \\ \cline{3-3}
                 &                & Atomic Comp. ($\%$)& 68.1 &  ---   &  ---  &  ---  &  ---    & 31.9  \\ \cline{2-9}
                 & Nb             & Binding Energy (eV)& 202.3 &  ---   &  ---  &  ---  &  ---   & ---    \\ \cline{3-3}
                 & Film           & FWHM (eV)          & 0.6  &  ---   &  ---  &  ---  &  ---   & ---    \\ \cline{3-3}
                 &                & Atomic Comp. ($\%$)& 100   &  ---   &  ---  &  ---  &  ---   & ---\\ \hline
    
    \end{tabular}
    
    \caption{\textbf{XPS Parameters Table for Nb2-4 and Nb Thin Film analysis}: A table showing the peak positions and fitting parameters used in the surface scan of Nb2-4. These fit values were used in the XPS peak fitting shown in  Fig.~\ref{Fig3} (b) and Fig.~\ref{SFig 4} (c).}

    \label{Table 2}
\end{table}

\twocolumngrid

% \begin{figure*}[ht]
%    \centering
%     \includegraphics[width= 0.75\textwidth]{Figures/SFig_5_Rev_2.pdf}
%     \caption{\textbf{XPS Parameters Table for Nb2-4 and Nb Thin Film analysis}: A table showing the peak positions and fitting parameters used in the surface scan of Nb2-4. These fit values were used in the XPS peak fitting shown in Fig.~\ref{Fig3} (b) and Fig.~\ref{SFig 4} (c).
%     }
%   	\label{SFig 5}
% \end{figure*} 

A \SI{5}{\kilo e \volt} monatomic argon ion sputter system was used to etch the surface \textit{in-situ}, allowing for depth profiling of the oxide thickness and characterization of deleterious compounds near the surface. To estimate the etch rate of the argon-ion sputter system, an Nb thin-film sample, with a well-characterized thickness was used. The Nb thin film was confirmed to be $73$~nm thick by Atomic Force Microscopy (AFM) on a Bruker Dimension Icon AFM instrument. The argon ion sputter time was 3 min per cycle, and after 11 cycles the Al peaks from the sapphire substrate became dominant, setting a nominal etch rate of $0.3-$\SI{0.4}{\angstrom/\sec}. For the data presented in Fig.~\ref{SFig 6}, a fine sputter of \SI{15}{s} per cycle was used. Results of the XPS on the Nb thin film control sample are presented in Fig.~\ref{SFig 4} (c). This sample, which was etched using a CF\textsubscript{4} plasma, and exposed to air for $>1$ month, shows a surface oxide composition similar to that of the Nb2-4 cavity shown in Fig.~\ref{Fig3} (b) after exposure for more than 3 weeks. Bulk analysis of the thin film however shows a deviation from the measured cavity samples. The Nb 3d peaks are far more resolved, with no sign of NbH\textsubscript{x} in the fitted spectra, further highlighting the effects of wet etching on the presence of hydrogen impurities in the bulk. 

\section{Nb oxide thickness analysis}

\label{section:oxide_thickness}
Two methods were used to determine the thickness of the niobium oxide layer. The first, "direct" method, was to directly measure Nb-oxide peaks for various sputter times and extract an approximate thickness based on the depletion of oxygen, while the second used relative peak intensities to indirectly determine the oxide thickness. For indirect thickness estimation, the peak intensities ($I_0/I_m$) ratio of the Nb 3d oxide phases to the Nb metal phase is used to calculate the oxide thickness at the surface. The surface oxide thickness, $d_{xps}$(nm), was estimated using the equation \cite{Alexander2002, Grundner1980, antoine1999morphological}.

\begin{equation}
    d_{xps}(nm)=\lambda\sin\theta\ln{\bigg(\frac{N_{m}\lambda _{m}I_{o}}{N_{o}\lambda_{o}I_{m}}+1\bigg)}
    \label{eqn:oxide thickness}
\end{equation}

\begin{figure}[!ht]

    \includegraphics[width= 0.4\textwidth]{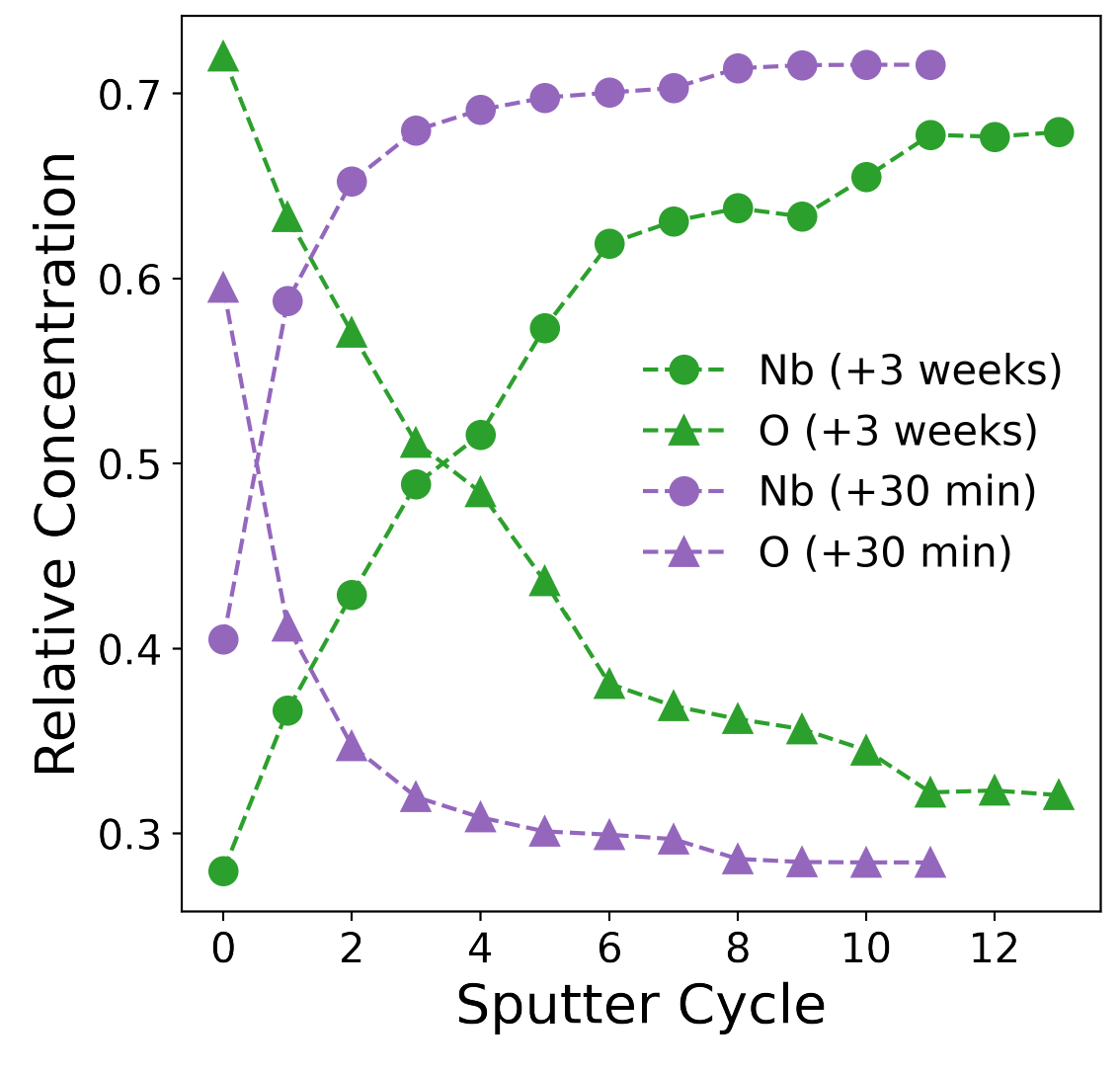}
    \caption{\textbf{Depth profile and extracted oxide thickness}: A plot of niobium metal and oxygen concentration based on XPS analysis. Each sputter cycle is 15 seconds, giving a cumulative etch of \SI{0.47}{nm} per sputter cycle. Longer air exposure requires more sputter cycles before reaching a stable value, indicating a thicker oxide. 
    }
  	\label{SFig 6}
\end{figure}

Where the ratio of the volume densities of Nb atoms in metal to different oxide phases ($N_m/N_o$) are listed in Table.~\ref{Table 2}. The inelastic mean free path (IMFP) values, $\lambda$, for Nb\textsubscript{2}O\textsubscript{5}, NbO\textsubscript{2}, and NbO can be found in Ref. \cite{NIST82}. These values are specified normal to the surface ($\theta$= 90$^\circ$) of the Kratos Axis Nova XPS instrument, which has an angle of 45$^\circ$ between the X-ray source and the axis of the electron lens. For direct determination, a fine sputter etch (15 s per cycle) was used, with the cycle repeated until the relative concentrations of both niobium and oxygen reached a constant value. Fig.~\ref{SFig 6} shows the niobium and oxygen concentrations from the survey scans as a function of sputter cycle. By 11 cycles the 3 week air exposed cavity show stable concentrations of oxygen and niobium, while the cavity exposed to air for 30 minutes reaches a stable value in 8 sputter cycles. Using the calibration in Appendix. ~\ref{section:surface_char}, this gives a directly measured oxide thickness of \SI{5.2}{nm} and \SI{3.8} {nm} for 3 week and 30 minute air exposure respectively.  The Nb oxide thickness extracted via the indirect method was determined to be \SI{4.9} {nm} and \SI{3.6}{nm} for 3 weeks and 30 minutes respectively. The estimated oxide thickness from the Nb thin film used for sputter calibration in Appendix. ~\ref{section:surface_char}, and Nb cavities with different air exposure duration are shown in Table \ref{Table 3}, along with the relative thicknesses of the three Nb oxidation states (NbO, NbO\textsubscript{2}, and Nb\textsubscript{2}O\textsubscript{5}).

\begin{table}[!h]
\begin{tabular}{|l|c|c|c|}
    \hline
      \textbf{Species} & \multicolumn{1}{|p{1.8cm}|}{\centering \textbf{Nb 75nm Thin Film}} & \multicolumn{1}{|p{1.75cm}|}{\centering \textbf{Nb 2-4}\\ \textbf{(+3 wks)}} & \multicolumn{1}{|p{1.75cm}|}{\centering \textbf{Nb 2-4}\\ \textbf{(+30 min)}} \\
    
      \hline
    %   \rowcolor{Rectcolor}
      Nb\textsubscript{2}O\textsubscript{5} & 4.5$\pm$0.01 nm & 4.4$\pm$0.01 nm & 2.4$\pm$0.04 nm \\ % <--
    %   \rowcolor{Rectcolor}
      NbO\textsubscript{2} & 2.1$\pm$0.02 nm & 0.4$\pm$0.01 nm &  0.5$\pm$0.02 nm \\ % <--
    %   \rowcolor{Rectcolor}
      NbO & 0.5$\pm$0.02 nm & 0.1$\pm$0.03 nm &  0.6$\pm$0.01 nm \\ % <--
      \hline
      Total & 7.1$\pm$0.02 nm & 4.9$\pm$0.01 nm &  3.5$\pm$0.01 nm \\ % <--
    %   \hline
    
      \hline
    \end{tabular}
    \caption{\textbf{Table of Extracted Oxide Thickness} A table summarizing the relative thickness of various oxide species using Eq.~\ref{eqn:oxide thickness}. The calculated values predict a systematically lower value when compared with the direct observation, however the difference is bound by the precision of the depth profiling shown in Fig.~\ref{SFig 6}, with a maximum depth resolution of \SI{0.47}{nm}.}
    \label{Table 3}
\end{table}

In both the direct and indirect measurement of oxide thickness, the total change corresponded to an $\approx$38$\%$ increase in thickness. The effective loss tangent product $F_e\tan{\delta_{TLS}}$ is proportional to thickness. The fit of the loss tangent data, shown in Fig.~\ref{Fig3} suggests a 75$\%$ increase, suggesting that the relative composition plays an equally important roll in enhancing dielectric loss. The indirect fit method shows that the relative thickness of the Nb\textsubscript{2}O\textsubscript{5} grew by $\approx88\%$ while the NbO\textsubscript{2} and NbO species decreased in relative thickness. This may suggest that Nb\textsubscript{2}O\textsubscript{5} may play an outsize role in harboring two-level systems. It may also suggest that slow oxide evolution in the presence of air may yield an oxide layer with defects capable of harboring two-level systems. In either case, further investigation would be required.

\bibliography{references}

\end{document}